\documentclass{article}

\usepackage{arxiv}

\usepackage[utf8]{inputenc} 
\usepackage[T1]{fontenc}    
\usepackage{hyperref}       
\usepackage{url}            
\usepackage{amsfonts}       
\usepackage{nicefrac}       
\usepackage{microtype}      
\usepackage{graphicx}
\usepackage{natbib}
\usepackage{doi}
\usepackage{amsmath}
\usepackage{natbib}         
\usepackage{booktabs}        
\usepackage{enumitem}
\usepackage{multirow}
\usepackage{threeparttable}
 \usepackage[normalem]{ulem}
 \useunder{\uline}{\ul}{}

\title{Bayesian Hierarchical Adaptive Random Partition (BHARP) Model for Factorial Adaptive Platform Trials}

\author{ 
\textbf{Xianglin Zhao}\\
Department of Epidemiology, Biostatistics, and Occupational Health\\
McGill University, Montreal, Quebec, Canada\\
\texttt{xianglin.zhao@mail.mcgill.ca}\\
\And
\textbf{Shirin Golchi}\\
Department of Epidemiology, Biostatistics, and Occupational Health\\
McGill University, Montreal, Quebec, Canada
}

\date{November 2026\\Preprint.}

\renewcommand{\shorttitle}{BHARP Model for Factorial Adaptive Platform Trials}

\hypersetup{
  pdftitle={Bayesian Hierarchical Adaptive Random Partition (BHARP) Model for Factorial Adaptive Platform Trials},
  pdfauthor={Xianglin Zhao, Shirin Golchi},
  pdfsubject={Statistics - Applications (stat.AP)}, 
  pdfkeywords={Adaptive platform trials;
Bayesian hierarchical models;
factorial designs; 
information borrowing; 
model uncertainty.}
}

\begin{document}
\maketitle

\begin{abstract}
    
Multi-factorial adaptive platform trial designs allow multiple interventions to be evaluated in combination, but the large number of regimen-stratum cells poses challenges for information borrowing. Prespecified borrowing structures might be mis-specified when the pattern of treatment effect heterogeneity is unknown, potentially affecting both estimation and adaptive decisions. We formalize a Bayesian adaptive decision framework with combined therapies across domains, multiple patient strata, intervention deactivation and response-adaptive randomization. Within this framework, we investigate the Bayesian Hierarchical Adaptive Random Partition (BHARP) model, which learns the borrowing structure from the data while accounting for model uncertainty. Through trial simulations, we compare BHARP with a prespecified borrowing structure under different treatment effect patterns and stratum prevalence ratios. BHARP generally achieves lower estimation error and fewer incorrect deactivation decisions in homogeneous and non-additive settings. When all the interventions are equivalent, BHARP effectively controls spurious intervention deactivation. This work provides a formalized and reproducible framework for multi-factorial adaptive platform trials and demonstrates that BHARP offers a flexible alternative to prespecified borrowing models. These findings highlight the trade-off between estimation efficiency and robustness to model mis-specification and underscore the importance of simulation-based calibration of adaptive decision rules.
\end{abstract}

\keywords{
Adaptive platform trials;
Bayesian hierarchical models;
factorial designs; 
information borrowing; 
model uncertainty
}

\section{Introduction}
\label{intro}

Adaptive platform trials have gained popularity across a wide range of disease areas, particularly during the COVID-19 pandemic \citep{pitre2023}, because of their potential to generate evidence efficiently and to continuously improve clinical practice \citep{angus2019}. In these trials, multiple treatments can be investigated within a continuing platform, with adjustments made in response to accumulating data  \citep{woodcock2017}. Adaptive decisions, such as the entry or removal of treatments, modifications to randomization schemes and early termination of the entire trial, are governed by prespecified rules, which are calibrated through trial simulations at the design stage \citep{chow2008,angus2019}.

While factorial designs have a long history originating in agricultural experiments, they are particularly attractive in the context of platform trials because they enable multiple treatment comparisons to be conducted simultaneously and interventions to be evaluated in combination \citep{white2022}, thereby more closely reflecting routine clinical practice \citep{lorenzi2025APTstroke}. 
A prominent example of a factorial adaptive platform trial is REMAP-CAP (Randomized Embedded Multifactorial Adaptive Platform for Community-Acquired Pneumonia) \citep{remapcap}, which combines interventions across multiple clinical domains into regimens. Similar domain-based multifactorial structures have also been adopted in the design of other adaptive platform trials \citep{incept2025, APTrise2024, lorenzi2025APTstroke, remapild2024,SNAP2023}. 

To clarify the key design features and their connection to the underlying statistical methodology, we draw a correspondence between REMAP-CAP terminology and standard statistical terminology.
In REMAP-CAP, a \textit{domain} denotes a set of competing, mutually exclusive interventions targeting a common disease mechanism; an \textit{intervention} is a specific treatment to be investigated in a domain; and a \textit{regimen} is a combination of assigned interventions across all domains. The corresponding statistical design-of-experiments terms for domain, intervention, and regimen are factor, level, and experimental condition, respectively.
\textit{Strata} are defined to accommodate potential treatment effect heterogeneity (TEH) across clinically relevant patient subgroups, and patients within each stratum are randomized to receive regimens. In this sense, strata play a role analogous to blocks in a blocked factorial design.
We retain the REMAP-CAP terminology throughout the paper while noting its correspondence to standard factorial design terminology. In addition, we refer to each regimen-stratum combination as a cell. 

Information borrowing across regimen-stratum cells can be particularly valuable when their treatment effects are sufficiently similar \citep{berry2013bhm}. Given the nature of factorial designs, even with a modest number of domains, interventions and strata, the number of cells can grow rapidly. Furthermore, information in some cells might be sparse due to unequal stratum prevalence and adaptive decisions, such as response-adaptive randomization and termination of recruitment in certain strata.

Various Bayesian models that accommodate information borrowing across patient subgroups may be extended to this setting \citep{zhouji2024review}. Conventional Bayesian hierarchical models (BHMs) assume full exchangeability across subgroup-specific parameters, allowing information borrowing through shrinkage toward a common mean \citep{thall2003, berry2013bhm}. Several extensions have been proposed to adapt the strength as well as the scope of information borrowing. For example, some approaches empirically calibrate the borrowing strength under the full exchangeability assumption \citep{cbhm2018}, while others relax this assumption to adjust the scope of information borrowing \citep{exnex2016, chenLee2019BACIS, JiangYuan2021, kang2021,zheng2022borrowing}.

More generally, the scope of borrowing can be determined based on the similarity among subgroup-specific parameters. A natural way to represent such a borrowing structure is through a partition of the subgroups into clusters, as adopted in several existing approaches \citep{geng2020,lin2021,luLee2023, zhao2023,chenLee2020,liu2022,ChuYuan2018blast,Psioda2021,chapple2018,asano2020}.  In the context of factorial experiments, partitioning the main effects and interactions has also been considered \citep{nobilegreen2000}.
In the multi-factorial adaptive trial setting, it is particularly useful to represent the borrowing structure as a partition obtained by clustering the cells, and cell-specific parameters assigned to the same cluster borrow information from one another. At one extreme, an independent model with no borrowing corresponds to a partition consisting of singleton clusters, whereas at the other extreme, a single cluster containing all cells allows information borrowing across all cells. Between these extremes lies a spectrum of intermediate structures that borrow information within particular subsets of cells. As the number of cells increases, the number of possible partitions grows rapidly, giving rise to a large partition space. 

In REMAP-CAP, the information-borrowing structure is governed by prespecified modeling assumptions and can be viewed as a combination of three local borrowing strategies. Such prespecification can be efficient when it aligns well with the underlying TEH pattern, but it does not account for model uncertainty about alternative borrowing structures and may result in inappropriate borrowing when the model is mis-specified. In an adaptive trial, such mis-specification can affect not only the estimation, but also the subsequent adaptive decisions.

We consider adaptive trial designs that incorporate data-adaptive borrowing models to learn plausible partitions from accumulating trial data.
In a Bayesian framework, the relative evidence for candidate partitions can be represented by a posterior distribution over the partition space. In practice, several partitions may remain plausible given the observed data. Therefore, uncertainty about the pattern of TEH translates into uncertainty about the appropriate borrowing structure, or more generally, model uncertainty. Ideally, such model uncertainty should be propagated into posterior inference \citep{draper1995, hoeting1999}.

When the REMAP-CAP design features are considered beyond the original clinical setting, a data-adaptive borrowing model may be particularly useful. We consider the recently proposed Bayesian hierarchical adaptive random partition (BHARP) model \citep{BHARP}, which accounts for model structure uncertainty by enabling posterior inference over the partition space.
BHARP is well suited to multi-factorial adaptive trial designs for the following reasons.
It provides a computationally efficient implementation that facilitates simulation-based evaluation and calibration of trial designs, which typically require repeated model fitting across multiple data-generating scenarios, decision rules and candidate thresholds. This is particularly relevant in multi-factorial designs involving multiple domains and strata, since the number of regimen–stratum cells can increase rapidly.

In addition, BHARP provides a generalizable modeling framework that requires relatively little modification across trial configurations. Because regimen–stratum cells are treated within a common partitioning framework and their similarity structure is learned from the data, the same implementation can be applied across a range of trial configurations and data-generating mechanisms. In contrast, prespecified models may be application-specific, as changes in the number of domains, intervention levels and strata can require substantial changes to the model parameterization and borrowing structure.  
These practical advantages make BHARP useful not only as an analysis model but also as a flexible reference model for evaluating adaptive decision rules in platform trials. 

In this work, we formalize the key design features of REMAP-CAP into a generalizable framework for factorial platform trials, including multiple intervention domains, patient strata, intervention deactivation, and response-adaptive randomization. Within this framework, we investigate BHARP as a data-adaptive alternative to prespecified borrowing structures. We conduct trial simulations to assess estimation performance and operating characteristics under a range of treatment effect patterns and stratum prevalence scenarios. Particular attention is given to how model mis-specification and unaccounted uncertainty about the borrowing structure affect estimation and subsequent adaptive decisions.

The remainder of this paper is organized as follows. 
Section~\ref{s:M} introduces the BHARP model and formalizes the REMAP-CAP decision rules within a factorial design framework. 
Section~\ref{s:Sim} evaluates the framework through trial simulations in a simplified trial setting and compares BHARP with a Bayesian hierarchical model with a prespecified borrowing structure. 
Finally, Section~\ref{s:Dis} discusses the implications, limitations, and potential extensions of this work.

\section{Methodology}
\label{s:M}
\subsection{Model}
\label{s:M_model}

Consider a factorial platform trial investigating $D$ treatment domains (or factors). For domain $d$, let $\mathcal V_d=\{1,\ldots,L_d\}$ denote the set of investigated competing interventions (or levels). In a complete factorial design, selecting one intervention from each domain yields $I=\prod_{d=1}^D L_d$ regimens. We represent the regimen composition by a design matrix
\[
\boldsymbol{X}=(x_{id})_{I \times D},\quad x_{id}\in \mathcal V_d
\]
where $x_{id}$ denotes the intervention assigned to regimen $i$ in domain $d$, and the $i$th row uniquely specifies the composition of regimen $i$.

Suppose the trial investigates $I$ regimens among $K$ patient strata. Let ${Y}_{ikn}$ denote the outcome of patient $n$ in stratum $k$ randomized to regimen $i$, corresponding to cell $(i,k)$. We assume
\[
{Y}_{ikn} \sim N(\theta_{ik},  \varsigma^{-1}),\qquad i=1,\dots,I; \quad k=1,\ldots,K;\quad n=1,\dots,N_{ik}
\]
where $N_{ik}$ is the number of patients in cell $(i,k)$,  $\theta_{ik}$ denotes the corresponding mean outcome, and $\varsigma$ is the common within-cell precision with the prior $\varsigma \sim \text{Gam}(a_{\text{cell}}, b_{\text{cell}})$. 

BHARP models the cell means $\{\theta_{ik}: i=1,\dots,I,\;k=1,\dots,K\}$ through a finite mixture model
\begin{gather*}
\theta_{ik} \mid
\boldsymbol{\mu},\boldsymbol{\sigma}, z_{ik} \sim  N(\mu_{z_{ik}},{\sigma_{z_{ik}}})\\
\tau \sim \text{Gam}(a_{\text{between}}, b_{\text{between}}), 
\quad \mu_t \mid \tau \sim N(0,\tau^{-1}), \quad
\sigma_t \sim \text{InvGam}(a_{\text{within}}, b_{\text{within}})
\end{gather*}
where \(\boldsymbol{\mu}=(\mu_1,\ldots,\mu_q)\) and \(\boldsymbol{\sigma}=(\sigma_1,\ldots,\sigma_q)\) denote the vectors of component means and variances, respectively, and  
\(\tau\) controls the prior dispersion of the component means. Specifically, each $\theta_{ik}$ corresponds to an allocation index $z_{ik}\in\{1,\dots,q\}$ assigning the cell mean $\theta_{ik}$ to one of $q$ mixture components with probabilities $(w_1,\ldots,w_q)$:
\[z_{ik} \mid q,\boldsymbol w  \sim \text{Categorical}(w_1,\ldots,w_q),\] where component weights $\boldsymbol w$
are assigned a symmetric Dirichlet prior, $Dirichlet(1,\ldots,1)$, and the number of mixture components, $q$, is a random variable with a prior distribution of the form $P(q) \propto (q)^\alpha, q=1,\ldots,Q_{\max}$. The allocation indices induce a partition of the regimen–stratum cells that determines the information-borrowing structure: cell means assigned to the same cluster borrow information through their shared distribution component. To promote appropriate information borrowing, the prior distributions are specified such that within-component variances are concentrated on small values, encouraging homogeneity among co-clustered cells and forming more clusters; larger numbers of mixture components are favored to reduce the risk of excessive borrowing; and  component means are regularized toward plausible values on the outcome scale, as implausible component locations could otherwise lead to inappropriate clustering structures. The borrowing structure is thus determined by inferring the latent allocation indices. BHARP uses reversible-jump MCMC to sample the mixture configurations and model parameters jointly, allowing posterior inference on the cell means to average over possible partitions and account for uncertainty in the borrowing structure.

\begin{figure}[htbp]
    \centering
\includegraphics[width=\textwidth]{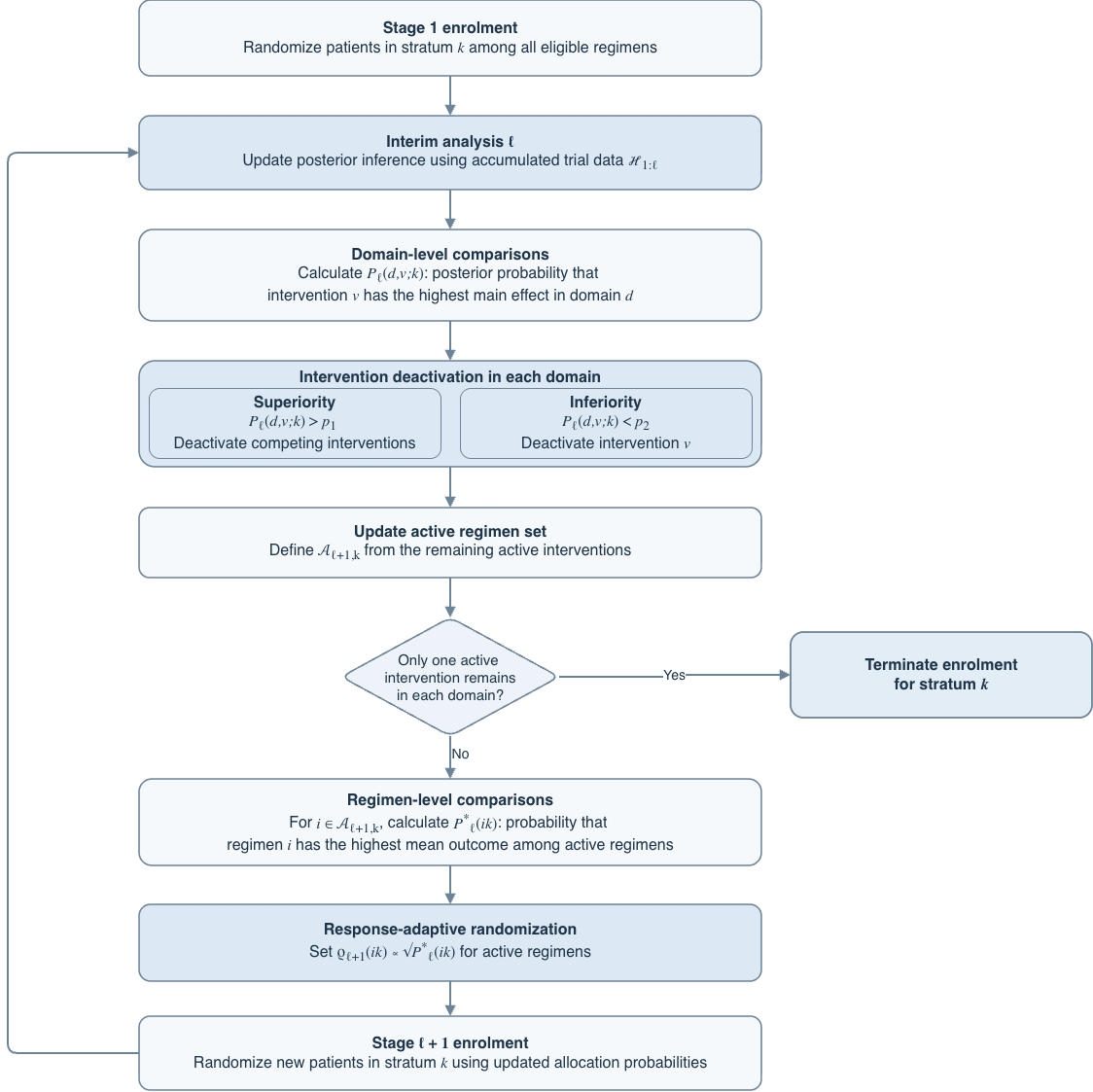}
    \caption{Adaptive decision process at each interim analysis for a given patient stratum \(k\). The decision process is repeated for each active stratum. The trial ends when all strata have terminated enrolment or the total sample size reaches \(N_{\max}\).}
    \label{fig:adaptive_design}
\end{figure}

\subsection{Adaptive Platform Trial Design}
\label{sec:decision_rules}
In this section, we specify a Bayesian adaptive platform trial design that uses posterior probabilities to guide intervention deactivation and response-adaptive randomization within each patient stratum. The design is motivated by the adaptive features of the REMAP-CAP platform trial.

To compare competing interventions within a treatment domain, we define main effects using marginal mean contrasts, as commonly defined in factorial experiments. Let $\mathcal I_{d,v}=\{i\in\{1,\dots,I\}: x_{id}=v\}$ denote the set of regimens containing intervention $v$ from domain $d$.
Let $\bar{\theta}_{\cdot k}$
denote the mean outcome across all regimens in stratum $k$ and $\bar{\theta}_{d,v;k}$ denote
the mean outcome among regimens containing intervention $v$ from domain $d$. The main effect of intervention $v$ in domain $d$ for stratum $k$ is then defined as
\[
\lambda_{d,v;k}=\bar{\theta}_{d,v;k}-\bar{\theta}_{\cdot k},\quad
 \bar{\theta}_{d,v;k}=
\frac{1}{\left| \mathcal I_{d,v}\right|}
\sum_{i\in\mathcal I_{d,v}}\theta_{ik},\quad
 \bar{\theta}_{\cdot k}=\frac{1}{I}\sum_{i=1}^{I}\theta_{ik}
.
\]

At each interim analysis \(\ell\), let \(\mathcal H_{1:\ell}\) denote the observed trial data accumulated up to interim analysis \(\ell\). Conditional on \(\mathcal H_{1:\ell}\), we use two types of posterior probabilities to guide adaptive decisions within each stratum \(k\).

\noindent\textbf{Comparing competing interventions within a domain.}
For stratum $k$, we calculate the posterior probability that intervention $v \in \mathcal V_d$ achieves the highest main effect in domain $d$ at interim analysis $\ell$:
\[
P_{\ell}(d,v;k)
=
P\!\left(
{\lambda}_{d,v;k}
=
\max_{v'\in\mathcal{V}_{d}}{\lambda}_{d,v';k}
\,\middle|\,\mathcal{H}_{1:\ell}
\right).
\]
Equivalently, this probability may be computed using $\bar{\theta}_{d,v;k}$, since $\bar{\theta}_{\cdot k}$ is common to all interventions in $\mathcal V_d$.

For each domain $d$, interventions are deactivated according to the following rules:
\[
\begin{aligned}
&\text{Superiority:} && P_{\ell}(d,v;k) > p_1 
    &&\Rightarrow \text{deactivate  competing levels } \mathcal{V}_d \setminus \{v\}, \\
&\text{Inferiority:} && P_{\ell}(d,v;k) < p_2 
    &&\Rightarrow \text{deactivate level } \{v\}.
\end{aligned}
\]
Once deactivated in a stratum, an intervention remains inactive for the remainder of the trial and is excluded from subsequent deactivation decisions.

\noindent\textbf{Response-adaptive randomization based on regimen comparisons.} Let \(\mathcal A_{\ell+1,k}\) denote the set of regimens remaining active in stratum \(k\) following the deactivation decisions at interim analysis \(\ell\). During stage \(\ell+1\), defined as the enrollment period between interim analyses \(\ell\) and \(\ell+1\), patients in stratum \(k\) are randomized among these active regimens.
For each active regimen $i \in A_{\ell+1,k}$, we calculate the posterior probability that regimen $i$ has the highest mean outcome among active regimens:

\[
P^*_{\ell}(ik) =
P\!\left(
\theta_{ik} 
=
\max_{i' \in \mathcal{A}_{\ell+1,k}}\theta_{i'k}
\,\middle|\,\mathcal{H}_{1:\ell}
\right).
\]
The randomization probabilities, \(\rho_{\ell+1}(ik)\) for the next stage are proportional to the square roots of these posterior probabilities:
\[
\rho_{\ell+1}(ik) =\frac{ \sqrt{P^*_{\ell}(ik)}}{\sum_{i' \in \mathcal{A}_{\ell+1,k}} \sqrt{P^*_{\ell}(i'k)}}
.
\]

Enrollment of a stratum is terminated once only one active intervention remains in each domain. The trial terminates when enrollment has stopped in all strata or when the total sample size reaches the prespecified maximum $N_{\max}$.

The overall adaptive decision process is summarized
in Figure~\ref{fig:adaptive_design}.

\section{Simulation Study}
\label{s:Sim}
In this section, we conduct simulation studies under a range of scenarios for a hypothetical adaptive clinical trial that investigates three domains with two, two and three available interventions each resulting in $I=12$ regimens. The corresponding design matrix is as follows

\begin{equation}
\label{eq:design_mat_223}
\boldsymbol X =
\begin{array}{c|ccc}
i & x_{i1} & x_{i2} & x_{i3} \\ \hline
1  & 1 & 1 & 1 \\
2  & 2 & 1 & 1 \\
3  & 1 & 2 & 1 \\
4  & 2 & 2 & 1 \\
5  & 1 & 1 & 2 \\
6  & 2 & 1 & 2 \\
7  & 1 & 2 & 2 \\
8  & 2 & 2 & 2 \\
9  & 1 & 1 & 3 \\
10 & 2 & 1 & 3 \\
11 & 1 & 2 & 3 \\
12 & 2 & 2 & 3
\end{array}
\end{equation}

Patients are classified into two prespecified strata, which may represent clinically distinct subgroups defined by biomarker or disease status. To investigate different assumptions of stratum prevalence, we evaluate the design under three stratum prevalence ratios: $1:1$, $1:5$ and $5:1$, representing a balanced population and two unbalanced populations in which one stratum accounts for approximately 17\% of randomized patients. Such imbalance may arise naturally when stratification is based on a relatively uncommon clinical characteristic or biomarker. 

Two interim analyses are conducted at cumulative total sample sizes of 600 and 900 patients, with a maximum total sample size of 1200. At each analysis, the decision rules described in Section~\ref{sec:decision_rules} are applied, using a superiority threshold of $p_1=0.95$ and an inferiority threshold of $p_2=0.05$. We also evaluate a more conservative specification with \(p_1=0.99, p_2=0.01\) to assess the sensitivity of the results to the decision thresholds.

For BHARP, we specify the hyperparameters as \((a_{\mathrm{cell}},b_{\mathrm{cell}})=(5,6)\), \((a_{\mathrm{between}},b_{\mathrm{between}})=(4,4)\), and \((a_{\mathrm{within}},b_{\mathrm{within}})=(120,0.03^2\times121)\). 
The induced prior on the observation variance has a mode of one, consistent with the unit variance assumed in the data-generating models. 
The prior on the within-component variance has a mode of \(0.03^2\), favoring information borrowing among regimen-stratum cells with similar treatment effects.
The prior on the component means allows relatively large between-component differences, while the prior on the number of components is specified as \(P(q)\propto q^2\), favoring more components to discourage excessive information borrowing. 
These prior specifications are held fixed across all simulation scenarios.

\subsection{Comparator model: prespecified borrowing structure}

We implement a simplified Bayesian hierarchical model motivated by the REMAP-CAP trial as a comparator. The original REMAP-CAP design is substantially more complex, employing Bayesian models for ordinal outcomes and incorporating additional covariates and interaction effects. Here, we consider normally distributed continuous outcomes, with unit observation variance in the data-generating model. Our objective is not to reproduce the original likelihood or prior distributions exactly, but to preserve the additive mean structure and representative hierarchical borrowing configurations. The additional complexities of the original model are not central to our comparison, which focuses on the impact of borrowing-structure assumptions on statistical inference and adaptive decisions. Since the simplified model retains the key borrowing features of the original formulation, our findings provide insights into the potential consequences of borrowing-structure mis-specification in complex models such as REMAP-CAP.

In the original REMAP-CAP formulation, the first intervention within each domain is treated as the reference level. The mean model is primarily additive in domain-specific intervention effects, while selected two-way interaction terms may also be included. Depending on prior clinical assumptions, these interactions may be constrained to zero or assigned normal priors, including strongly informative \(N(0,0.05^2)\) priors centered at zero on the log-odds scale.  
In our simplified comparator, we omit these terms while retaining the prespecified hierarchical borrowing structure. For the $2\times2\times3$ factorial design considered in this simulation study, the mean response for regimen $i$ in stratum $k$ is decomposed into a stratum-specific baseline effect and additive domain-specific intervention effects: 
\[
{Y}_{ikn} \sim N(\mu_k+\beta_{1,x_{i1};k}+\beta_{2,x_{i2};k}+\beta_{3,x_{i3};k},\ \varsigma^{-1}),\quad \beta_{1,1;k}=\beta_{2,1;k}=\beta_{3,1;k}=0,
\] 

The information-borrowing structure in the original REMAP-CAP model is intervention-specific and can be viewed as a combination of three local borrowing strategies. Specifically, different intervention effects are assigned different prior structures, including (a) no borrowing; (b) borrowing across two strata for the same intervention; and (c) borrowing across two interventions within each stratum at the first hierarchical level, followed by borrowing across strata at the second level.
In practice, such specifications are likely justified by clinical judgment or prior information and beliefs. Although the degree of borrowing can be adjusted through the posterior distributions of the hierarchical variance parameters, the grouping structure itself remains fixed. Consequently, mis-specification of the assumed borrowing structure may lead to inefficient borrowing, biased estimation, or inadequate uncertainty quantification.

To mimic this setting, we construct the REMAP-CAP-style comparator with a hypothetical borrowing structure prespecified based on clinical judgment. Specifically, the expert recommends (a) no borrowing across strata for domain 1, (b) borrowing across strata for the non-reference intervention effect in domain 2, and (c) joint borrowing across non-reference interventions and strata for domain 3. This comparator represents an expert-specified borrowing model, in contrast to a data-adaptive approach that learns the borrowing structure from the observed data while accounting for uncertainty in its specification. The hypothetical borrowing structure is fixed throughout the simulation study, allowing us to evaluate how a clinically specified REMAP-CAP-style comparator performs under different degrees of borrowing-structure mis-specification. Following the notation in REMAP-CAP, the borrowing model is specified as follows:

\noindent Domain 1:
\[\beta_{1,2;1} \sim N(0,3^2),\quad \beta_{1,2;2}\sim N(0,3^2).\]

\noindent Domain 2:
\begin{gather*}
    \beta_{2,2;1},\beta_{2,2;2} \sim N(\mu_{2,2},\sigma_{2,2}),\\
\mu_{2,2} \sim N(0,3^2),\quad \sigma_{2,2} \sim InvGam(0.125,0.00281)
\end{gather*}

\noindent Domain 3:
\begin{gather*}
\beta_{3,2;1}, \beta_{3,3;1} \sim N(\mu_{3,2+3;1},\tau_{3,2+3}), \\
\beta_{3,2;2},\beta_{3,3;2} \sim N(\mu_{3,2+3;2},\tau_{3,2+3}),\\
\mu_{3,2+3;1}, \mu_{3,2+3;2} \sim N(\mu_{3,2+3},\sigma_{3,2+3}), \\
\mu_{3,2+3} \sim N(0,3^2), \quad \sigma_{3,2+3} \sim InvGam(0.125,0.00281),\\
\quad \tau_{3,2+3} \sim InvGam(0.125,0.00281),
\end{gather*}
where the subscript 3,2+3 indicates that the corresponding parameter is shared by intervention 2 and 3 in domain 3. Throughout this specification, the second parameter of the normal distribution denotes the variance, and the inverse-gamma distributions are parameterized by shape and scale.

As in BHARP, the observation precision is assigned a \(\mathrm{Gam}(5,6)\) prior. 
The stratum-specific baseline means \(\mu_1,\mu_2\) are assigned independent \(N(0,1)\) priors, which is weakly informative relative to the effect sizes considered in the simulation study. The intervention effects in domain 1, \(\beta_{1,2;1},\beta_{1,2;2}\), and the top-level means of the hierarchical distributions in domains 2 and 3, \(\mu_{2,2},\mu_{3,2+3}\), are assigned independent \(N(0,3^2)\) priors, which is diffuse on the effect-size scale.
The hierarchical variance parameters are assigned independent inverse-gamma priors with shape parameter 0.125 and scale parameter 0.00281, using the hyperparameter values specified for nested intervention effects in an earlier REMAP-CAP statistical analysis plan. These variance parameters allow the degree of information borrowing to be estimated from the data, while the hierarchical borrowing structure remains prespecified. 

Since the comparator is parameterized in terms of intervention effects relative to the reference level, the posterior probability of superiority can be calculated directly from the corresponding $\beta_{d,v;k}$:
\[
P_{\ell}(d,v;k) = P\!\left( \beta_{d,v;k} = \max_{v'\in\mathcal{V}_{d}}\beta_{d,v';k}\,\middle|\,\mathcal{H}_{1:\ell}
\right),
\]
where the reference intervention is included in this comparison through the fixed value $\beta_{d,1;k}=0$. Under the additive model, the ranking of domain-specific intervention effects is equivalent to that of the corresponding marginal main effects. These posterior probabilities are then used in the same decision rules described in Section~\ref{sec:decision_rules}.

\subsection{Scenarios}

We consider four data-generating scenarios designed to evaluate the operating characteristics of the proposed model and the REMAP-CAP-style comparator. The numerical values used in these scenarios are chosen with reference to conventional standardized effect-size benchmarks. Because the outcome standard deviation is set to one in the simulation, mean differences of 0.2, 0.5, and 0.8 correspond to small, moderate, and large standardized effects, respectively, following Cohen's conventional guidelines \citep{cohen1988}.  The true mean responses, denoted by $\theta_{ik}$ for regimen $i$ in stratum $k$, are specified below. 

\noindent \textbf{Scenario S1}:

\[\theta_{ik}=0.2, \qquad i=1,\ldots,12,\quad k=1,2.\]

\noindent \textbf{Scenario S2}: 
\[
\theta_{ik} =
\begin{cases}
0.8, & \text{if } (i,k)\in\{(4,1),(11,1),(12,1),(10,2)\},\\
0, & \text{otherwise}.
\end{cases}
\]

\noindent \textbf{Scenario S3}: 
\begin{align*}
    \theta_{i1} &= 0.2 + 0.3\,\mathbf{1}(x_{i1}=2,\ x_{i3}=3) + 0.3\,\mathbf{1}(x_{i2}=2,\ x_{i3}=3), \\
    \theta_{i2} &= 0.6\,\mathbf{1}(x_{i2}=1,\ x_{i3}=3) + 0.2\,\mathbf{1}(x_{i1}=2).
\end{align*}

\noindent \textbf{Scenario S4}: 
\begin{align*}
    \theta_{i1} &= 0.5\,\mathbf{1}(x_{i3}=2),\\
    \theta_{i2} &= 0.5+0.3\,\mathbf{1}(x_{i2}=2).
\end{align*}

Scenario S1 represents a completely homogeneous setting with no treatment effect heterogeneity across regimens or strata, where all regimens yield the same mean response of 0.2 in both strata. This scenario assesses whether the methods avoid introducing spurious treatment effect heterogeneity and making unwarranted intervention deactivation decisions when the true effects are homogeneous.
Scenarios S2 and S3 represent non-additive settings in which the true regimen effects cannot be adequately described by main effects alone. In Scenario S2, nonzero mean responses are confined to a small number of regimens within each stratum, creating sparse cell-level signals. In stratum 1, regimen 4, 11, and 12 are effective, while in stratum 2, regimen 10 is the only effective regimen. The scenario also yields a unique optimal intervention level for each domain within each stratum: for stratum $k=1$, $(D1,v2)$, $(D2,v2)$ and $(D3,v3)$ are optimal; for stratum $k=2$, $(D1,v2)$, $(D2,v1)$ and $(D3,v3)$ are optimal. Scenario S3 generates treatment effects primarily through two-way interactions between domains. This scenario evaluates whether the methods can identify optimal interventions when the effect of an intervention depends on which intervention is assigned in another domain. Similar to Scenario S2, this scenario yields one optimal intervention level for each domain within each stratum. Despite their different underlying patterns of TEH, Scenarios S2 and S3 yield the same optimal intervention levels in both strata.
Scenario S4 provides a favorable setting for the REMAP-CAP-style model, where the true mean structure is additive in domain-level effects. Within each stratum, interventions are equivalent in two domains, while the remaining domain contains a unique optimal intervention: for stratum $k=1$, treatment effects vary only across interventions in domain 3, with level 2 being optimal, whereas for stratum $k=2$, treatment effects vary only across interventions in domain 2, with level 2 being optimal.  This scenario is included to examine whether the BHARP model remains competitive in a setting where the additive mean structure assumed by the comparator is correctly specified. The coexistence of domains with and without intervention-level heterogeneity also allows us to assess the reliability of adaptive decisions in mixed settings.

\subsection{Results}

For each scenario, 500 simulated trials are generated. To evaluate the estimation performance and the operating characteristics of the adaptive design, we assess the root mean squared error (RMSE) for each cell $(i,k)$, computed across simulated trials using the posterior median of \(\theta_{ik}\) as the point estimate. We also examine the empirical distribution of the trial termination stage. The main results are presented for the decision thresholds $p_1=0.95$ and $p_2=0.05$, with additional results under the more conservative thresholds $p_1=0.99$ and $p_2=0.01$.

To evaluate the domain-specific superior/inferior decisions within each stratum, we assess the probability of incorrectly deactivating a truly optimal intervention. For stratum $k$, we define the set of optimal interventions in domain $d$ as 
$$\mathcal{B}_{d;k}=\left\{ v\in\mathcal V_d: \bar\theta_{d,v;k} = \max\limits_{v'\in \mathcal V_d} \bar\theta_{d,v';k} \right\}.$$ Optimality is defined in terms of domain-specific mean responses, and an incorrect deactivation occurs if any intervention in $\mathcal{B}_{d;k}$ is deactivated, either following an inferiority conclusion for itself or a superiority conclusion for its competitor.  If all interventions in a domain have equal main effects, any superiority or inferiority conclusion is considered incorrect.

\begin{figure}[htbp]
  \centering
  \includegraphics[width=\textwidth]{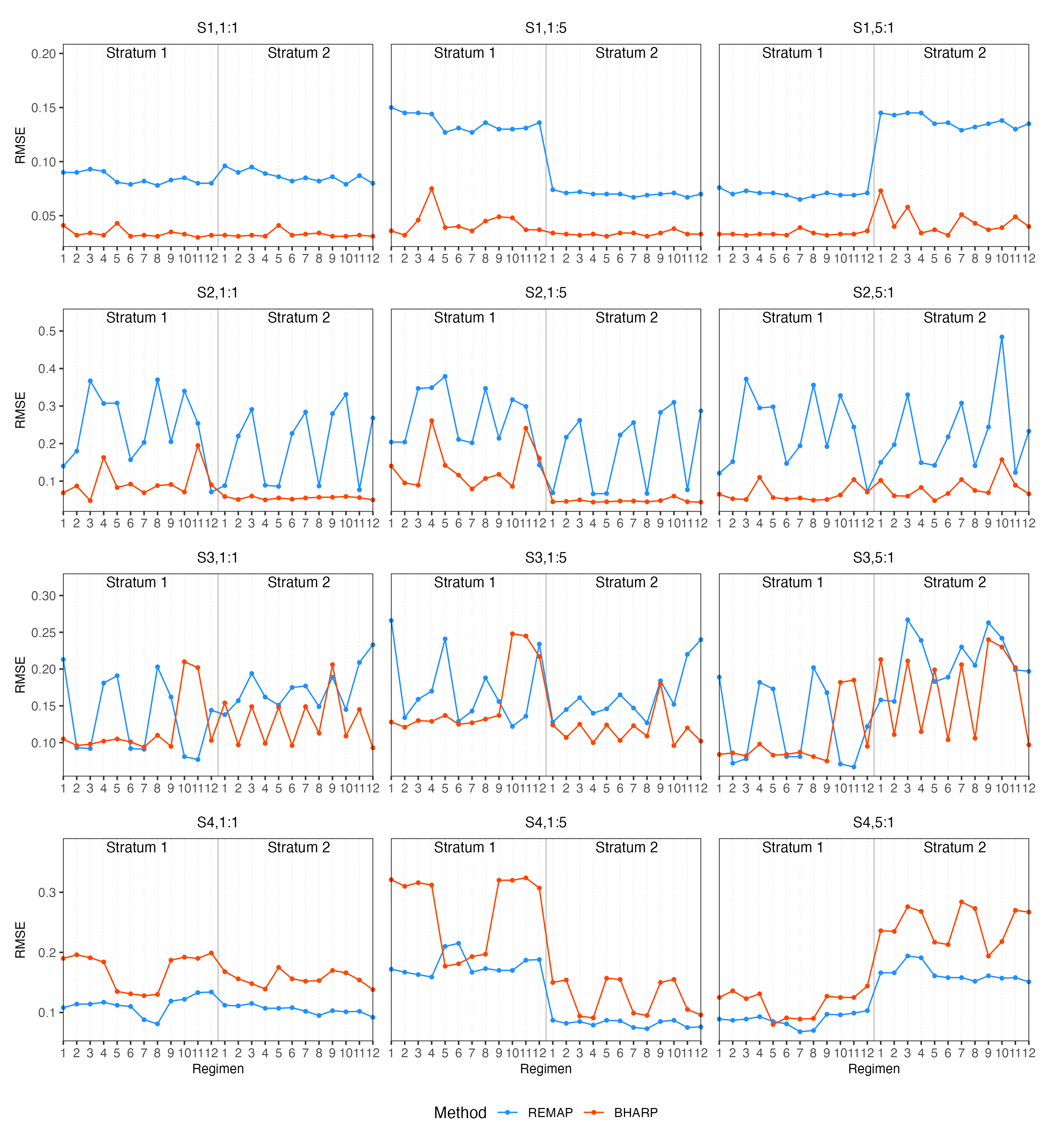}
  \caption{Root mean squared error (RMSE) of cell-mean estimates by scenario, stratum prevalence ratio, stratum, and regimen.}
  \label{fig:rmse_9505}
\end{figure}

\begin{figure}[htbp]
  \centering
  \includegraphics[width=\textwidth]{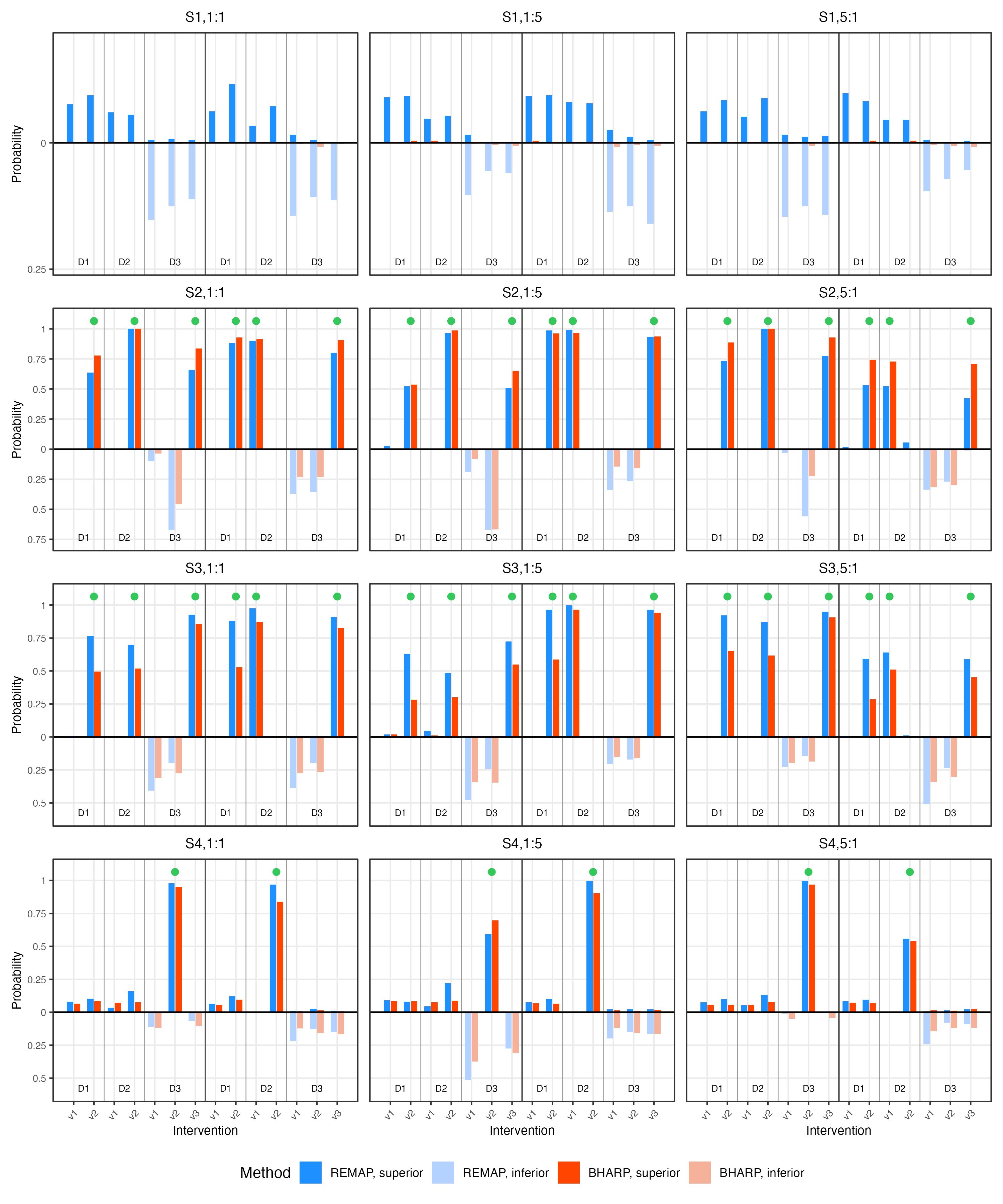}
  \caption{Empirical probabilities of superiority and inferiority declarations by scenario, stratum prevalence ratio, stratum, and domain. Green dots indicate uniquely optimal interventions.}
  \label{fig:dec_9505}
\end{figure}

\begin{table}[t]
\centering
\caption{Empirical distribution of trial termination across analysis stages by scenario and stratum prevalence ratio.}
\label{tab:stage_distribution_9505}
\small
\setlength{\tabcolsep}{4pt}
\begin{tabular}{lcccccc}
\toprule
&  \multicolumn{2}{c}{\textbf{Stage 1}} 
& \multicolumn{2}{c}{\textbf{Stage 2}} 
& \multicolumn{2}{c}{\textbf{Stage 3}} \\
\cmidrule(lr){2-3} \cmidrule(lr){4-5} \cmidrule(lr){6-7}
\textbf{Scenario} 
& \textbf{REMAP} & \textbf{BHARP} 
& \textbf{REMAP} & \textbf{BHARP} 
& \textbf{REMAP} & \textbf{BHARP} \\
\midrule
\textbf{S1, 1:1} & 0.00 & 0.00 & 0.00 & 0.00 & 1.00 & 1.00 \\
\textbf{S1, 1:5} & 0.00 & 0.00 & 0.00 & 0.00 & 1.00 & 1.00 \\
\textbf{S1, 5:1} & 0.00 & 0.00 & 0.00 & 0.00 & 1.00 & 1.00 \\
\midrule
\textbf{S2, 1:1} & 0.00 & 0.10 & 0.13 & 0.34 & 0.87 & 0.56 \\
\textbf{S2, 1:5} & 0.00 & 0.01 & 0.05 & 0.16 & 0.95 & 0.83 \\
\textbf{S2, 5:1} & 0.00 & 0.01 & 0.07 & 0.27 & 0.93 & 0.72 \\
\midrule
\textbf{S3, 1:1} & 0.00 & 0.00 & 0.12 & 0.03 & 0.88 & 0.97 \\
\textbf{S3, 1:5} & 0.00 & 0.00 & 0.05 & 0.02 & 0.95 & 0.98 \\
\textbf{S3, 5:1} & 0.00 & 0.00 & 0.03 & 0.02 & 0.97 & 0.98 \\
\midrule
\textbf{S4, 1:1} & 0.00 & 0.00 & 0.00 & 0.00 & 1.00 & 1.00 \\
\textbf{S4, 1:5} & 0.00 & 0.00 & 0.00 & 0.00 & 1.00 & 1.00 \\
\textbf{S4, 5:1} & 0.00 & 0.00 & 0.00 & 0.00 & 1.00 & 1.00 \\
\bottomrule
\end{tabular}
\end{table}

\begin{table}[t]
\centering
\caption{Empirical probability of incorrect intervention deactivation by scenario, stratum prevalence ratio, stratum, and domain.}
\label{tab:incorrect_decisions_9505}
\small
\setlength{\tabcolsep}{4pt}
\begin{tabular}{l l ccc ccc c}
\toprule
&  & \multicolumn{3}{c}{\textbf{Stratum 1}} 
& \multicolumn{3}{c}{\textbf{Stratum 2}} 
& \\
\cmidrule(lr){3-5} \cmidrule(lr){6-8}
Scenario 
& Method
& \textbf{D1} & \textbf{D2} & \textbf{D3}
& \textbf{D1} & \textbf{D2} & \textbf{D3}
& \textbf{Trial} \\
\midrule
\textbf{S1, 1:1} & 
REMAP & 0.17 & 0.12 & 0.39 & 0.18 & 0.11 & 0.36 & 0.78 \\
& BHARP & 0.00 & 0.00 & 0.00 & 0.00 & 0.00 & 0.01 & 0.01 \\
\textbf{S1, 1:5} & 
REMAP & 0.18 & 0.10 & 0.22 & 0.19 & 0.16 & 0.43 & 0.79 \\
& BHARP & 0.01 & 0.01 & 0.01 & 0.01 & 0.00 & 0.02 & 0.03 \\
\textbf{S1, 5:1} & 
REMAP & 0.15 & 0.14 & 0.41 & 0.18 & 0.09 & 0.21 & 0.76 \\
& BHARP & 0.00 & 0.00 & 0.01 & 0.01 & 0.00 & 0.02 & 0.03 \\
\midrule
\textbf{S2, 1:1} & 
REMAP & 0.01 & 0.00 & 0.00 & 0.00 & 0.01 & 0.00 & 0.02 \\
& BHARP & 0.00 & 0.00 & 0.00 & 0.00 & 0.00 & 0.00 & 0.00 \\
\textbf{S2, 1:5} & 
REMAP & 0.02 & 0.00 & 0.01 & 0.00 & 0.00 & 0.00 & 0.03 \\
& BHARP & 0.00 & 0.00 & 0.00 & 0.00 & 0.00 & 0.00 & 0.01 \\
\textbf{S2, 5:1} & 
REMAP & 0.01 & 0.00 & 0.00 & 0.02 & 0.06 & 0.01 & 0.08 \\
& BHARP & 0.00 & 0.00 & 0.00 & 0.00 & 0.00 & 0.00 & 0.01 \\
\midrule
\textbf{S3, 1:1} & 
REMAP & 0.01 & 0.00 & 0.00 & 0.00 & 0.00 & 0.00 & 0.01 \\
& BHARP & 0.00 & 0.00 & 0.00 & 0.00 & 0.00 & 0.00 & 0.01 \\
\textbf{S3, 1:5} & 
REMAP & 0.02 & 0.05 & 0.00 & 0.00 & 0.00 & 0.00 & 0.07 \\
& BHARP & 0.02 & 0.01 & 0.00 & 0.00 & 0.00 & 0.00 & 0.03 \\
\textbf{S3, 5:1} & 
REMAP & 0.00 & 0.00 & 0.00 & 0.01 & 0.01 & 0.00 & 0.03 \\
& BHARP & 0.00 & 0.00 & 0.00 & 0.00 & 0.00 & 0.00 & 0.01 \\
\midrule
\textbf{S4, 1:1} & 
REMAP & 0.18 & 0.19 & 0.00 & 0.19 & 0.00 & 0.49 & 0.72 \\
& BHARP & 0.15 & 0.15 & 0.00 & 0.15 & 0.00 & 0.45 & 0.63 \\
\textbf{S4, 1:5} & 
REMAP & 0.17 & 0.26 & 0.00 & 0.17 & 0.00 & 0.51 & 0.78 \\
& BHARP & 0.17 & 0.16 & 0.00 & 0.13 & 0.00 & 0.43 & 0.63 \\
\textbf{S4, 5:1} & 
REMAP & 0.17 & 0.18 & 0.00 & 0.18 & 0.00 & 0.41 & 0.70 \\
& BHARP & 0.11 & 0.13 & 0.00 & 0.14 & 0.00 & 0.39 & 0.59 \\

\bottomrule
\end{tabular}

\begin{minipage}{0.55\linewidth}
\vspace{4pt}
\footnotesize
\textit{Note:} Incorrect deactivation refers to deactivation of at least one true optimal intervention. The ``Trial'' column reports the probability of at least one incorrect deactivation anywhere in the trial.
\end{minipage}

\end{table}

Scenario 1 represents a homogeneous setting in which all regimens have the same mean response of 0.2. REMAP performs considerably worse than BHARP in terms of both estimation accuracy and adaptive decisions. Compared to that of BHARP, the RMSE of REMAP is higher, and more sensitive to imbalanced prevalence ratios: for the rare stratum, the RMSE of REMAP reaches around three times as high as that of BHARP (Figure~\ref{fig:rmse_9505}). Although both methods allow the entire trial to continue to the final analysis stage without early termination (Table~\ref{tab:stage_distribution_9505}), their decisions within domains differ substantially. Ideally, no superiority or inferiority conclusions should be made in any domain since all interventions show no difference. BHARP largely achieves this objective: across all the prevalence ratios, the probability of drawing any false conclusion is close to zero for all domains (Figure~\ref{fig:dec_9505}). In contrast, REMAP tends to draw spurious superiority conclusions in two-level domains and inferiority conclusions in domains with more than two interventions, despite the absence of true differences. As a result, the probability of incorrectly deactivating at least one intervention anywhere in the trial ranges from 76\% to 79\% for REMAP, compared with only 1\% to 3\% for BHARP (Table~\ref{tab:incorrect_decisions_9505}).

\begin{figure}[htbp]
  \centering
  \includegraphics[width=\textwidth]{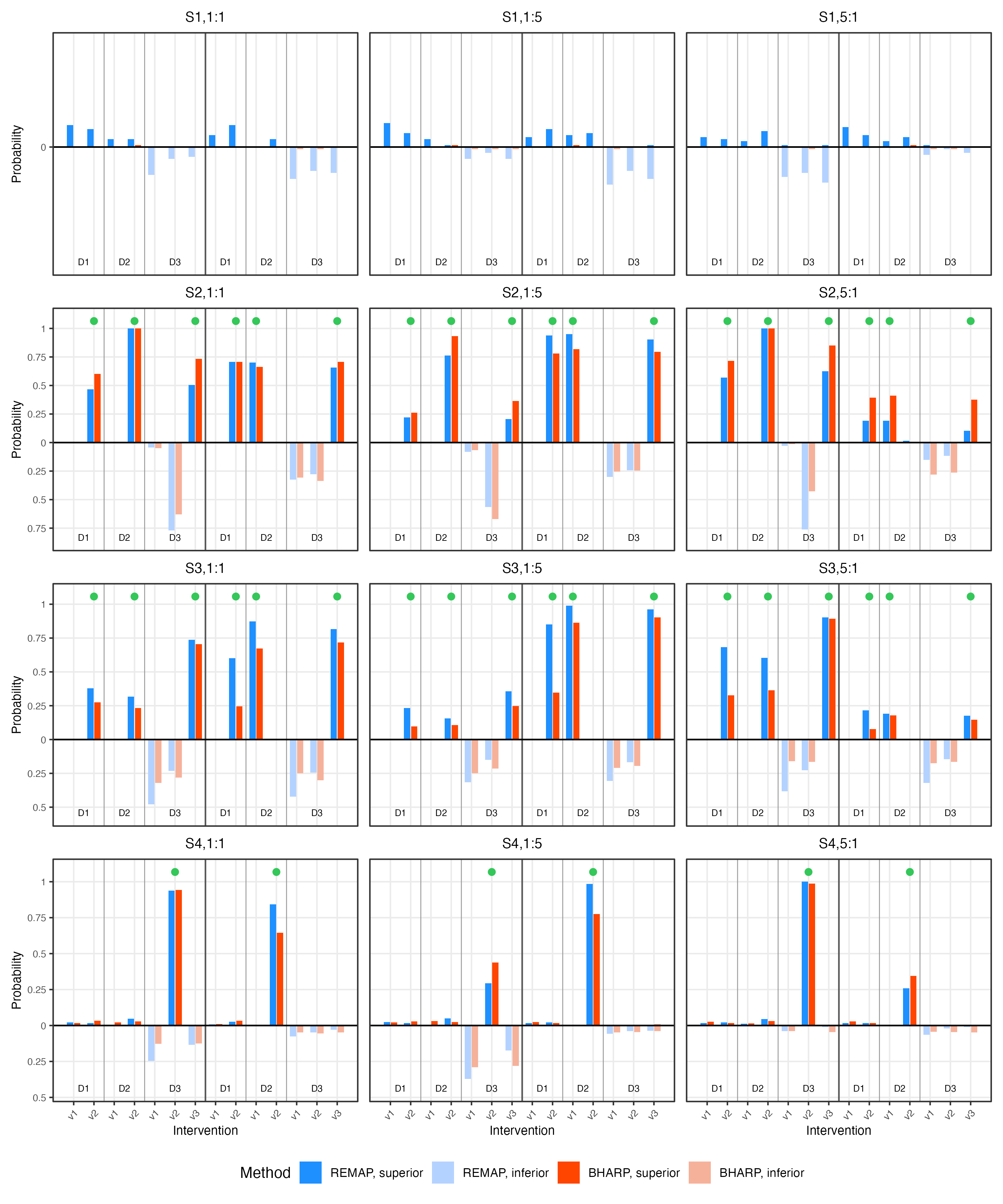}
  \caption{Empirical probabilities of superiority and inferiority declarations by scenario, stratum prevalence ratio, stratum, and domain under the conservative decision thresholds $p_1=0.99$ and $p_2=0.01$. Green dots indicate uniquely optimal interventions.}
  \label{fig:dec_9901}
\end{figure}

In Scenario 2, only regimens 4, 11, and 12 in stratum 1 and regimen 10 in stratum 2 have nonzero true mean responses. In both strata, each domain has a uniquely optimal intervention. BHARP generally has higher probabilities of correctly declaring superiority (Figure~\ref{fig:dec_9505}), even when more conservative decision thresholds are applied (Figure~\ref{fig:dec_9901}).
Table~\ref{tab:stage_distribution_9505} shows that REMAP rarely terminates trials early, whereas BHARP is more likely to terminate at the first or second interim analysis, particularly under balanced stratum prevalence. Consequently, a larger proportion of trials terminate with total sample sizes of 600 or 900 rather than reaching the maximum of 1,200.  However, despite the higher probability of early termination, BHARP achieves lower RMSE than REMAP in nearly all regimen-stratum cells (Figure~\ref{fig:rmse_9505}). Table~\ref{tab:incorrect_decisions_9505} further shows that BHARP maintains low probabilities of incorrect deactivation across all stratum prevalence ratios.

In Scenario 3, although the domain-specific optimal interventions are the same as in Scenario 2, the true effects are primarily driven by two-way interactions. Both methods have low probabilities of early termination (Table~\ref{tab:stage_distribution_9505}) and incorrect deactivation decisions (Table~\ref{tab:incorrect_decisions_9505}). However, the two methods exhibit different strengths: REMAP generally has higher probabilities of correctly declaring superiority (Figure~\ref{fig:dec_9505}), while BHARP generally achieves lower RMSE (Figure~\ref{fig:rmse_9505}) and lower probabilities of incorrect deactivation under imbalanced stratum prevalence. These results indicate that improved estimation accuracy does not necessarily translate into better performance on every adaptive decision metric.

In Scenario 4, the true effects follow an additive structure, which is correctly specified by REMAP. Figure~\ref{fig:rmse_9505} shows that REMAP achieves lower RMSE and is less sensitive to imbalanced stratum prevalence. For both strata, one domain has a uniquely optimal intervention while the other two domains are flat. Both methods show high probabilities of detecting the optimal interventions (Figure~\ref{fig:dec_9505}) without terminating the trial early (Table~\ref{tab:stage_distribution_9505}). However, both methods make spurious deactivation decisions in flat domains, where no intervention should be deactivated. Compared to BHARP, REMAP is slightly more prone to such incorrect deactivation decisions. Consequently, the trial-level probability of at least one incorrect deactivation is high, which is $70\%$--$78\%$ for REMAP and $59\%$--$63\%$ for BHARP. By applying more conservative decision thresholds, these probabilities are substantially reduced to 21\%–26\% for REMAP and 24\%–28\% for BHARP, resulting in comparable performance between the two methods (Table~\ref{tab:incorrect_decisions_9901}). However, the reduction in incorrect deactivation comes at the cost of lower probabilities of correctly declaring superiority, particularly in the less prevalent stratum (Figure~\ref{fig:dec_9901}).

\section{Discussion}
\label{s:Dis}

In this work, we formalized a generalizable Bayesian adaptive decision framework for factorial platform trials motivated by key design features of REMAP-CAP, including intervention domains, patient strata, within-domain intervention deactivation and response-adaptive randomization. This formulation provides a reproducible framework in which the trial decision rules, analysis models and information-borrowing strategies can be evaluated through trial simulation, facilitating the evaluation and adaptation of these design features to other settings. To illustrate the application of this framework, we designed scenarios with various data-generating mechanisms and stratum prevalence ratios. In this trial simulation, we investigated BHARP as a data-adaptive alternative to a prespecified borrowing model, allowing the borrowing structure across regimen-stratum cells to be learned from the observed data while accounting for the uncertainty in its specification. 

An important consideration in adaptive platform trials is that an improper borrowing structure can affect not only estimation but also subsequent trial decisions, which can sometimes have major consequences. For example, in the design, a false deactivation decision for an intervention can affect all combined regimens containing that intervention within the corresponding stratum. This is particularly evident in the homogeneous setting of Scenario 1. In this homogeneous setting, BHARP can adaptively learn the similarity structure across regimen-stratum cells and borrow information more extensively, reducing spurious differences in the estimated treatment effects. Therefore, when no true differences exist, BHARP is less likely to induce spurious contrasts and consequently less likely to trigger erroneous superiority or inferiority decisions. In contrast, even though the first interim analysis is conducted after half of the maximum sample size has been accrued, the prespecified model remains sensitive to random fluctuations, leading to erroneous decisions in approximately three quarters of the simulated trials.

These results illustrate the trade-off between efficiency under correct structural assumptions and robustness to model mis-specification. When the additive mean structure assumed by the REMAP-CAP-style model is correctly specified, more efficient estimation can be achieved. However, its reliance on prespecified assumptions becomes less advantageous when the true treatment effects deviate from the assumed additive structure. In contrast, BHARP may sacrifice some efficiency under correct specification in exchange for greater flexibility and robustness when the underlying treatment-effect structure is uncertain.

Our results highlight the importance of sensitivity analysis in assessing how decision thresholds affect the operating characteristics of the adaptive design. Under the primary thresholds ($p_1=0.95,p_2=0.05$), both methods show low probabilities of incorrect deactivation when each domain contains a unique optimal intervention, and BHARP further shows good control of spurious deactivation when all domains are flat. However, the current design may require additional calibration in mixed settings where domains with a unique optimal intervention coexist with flat domains: both methods frequently make incorrect deactivation decisions in flat domains. The sensitivity analysis shows that more conservative thresholds ($p_1=0.99,p_2=0.01$) substantially reduce incorrect deactivation in the flat domains of Scenario 4.

Several directions for extensions and future work arise from the present article. 
Future work could investigate how the proposed approach compares to prespecified modeling frameworks when additional sources of temporal complexity, such as time-varying patient states, are incorporated.  
Future simulation studies may be considered to evaluate the proposed approach in extended and/or incomplete factorial platforms and alternative outcome types. 
Finally, the simulation results indicate that the current decision framework can be vulnerable when domains with an optimal intervention coexist with flat domains. Beyond calibrating decision thresholds, further work could also investigate modifications to the decision rules to protect against spurious deactivation in such mixed settings.

Overall, this work provides a formalized and reproducible framework for multi-factorial adaptive platform trials motivated by the REMAP-CAP clinical trial. Within this framework, BHARP offers a flexible alternative to prespecified borrowing models by learning the information-sharing structure across regimen-stratum cells from the observed data while accounting for model uncertainty. Our simulation study illustrates the trade-off between estimation efficiency and robustness to structural assumptions. The sensitivity analysis further underscores the importance of calibrating the adaptive decision rules across heterogeneous scenarios. Together with its computational tractability and generalizable implementation, BHARP provides a practical tool for both the analysis and simulation-based development of complex adaptive platform trials.

\section*{Funding}

This research was supported by a doctoral scholarship from the Fonds de recherche du Québec -- Nature et technologies (FRQNT). SG acknowledges support from the Natural Sciences and Engineering Research Council of Canada (NSERC), Canadian Institute for Statistical Sciences (CANSSI), Fonds de recherche du Québec - Santé (FRQS) and FRQNT.

\section*{Conflicts of Interest} 
The authors declare no conflicts of interest.

\bibliographystyle{unsrtnat}
\bibliography{cite}  

\newpage
\begin{figure}[htbp]
  \centering
  \includegraphics[width=\textwidth]{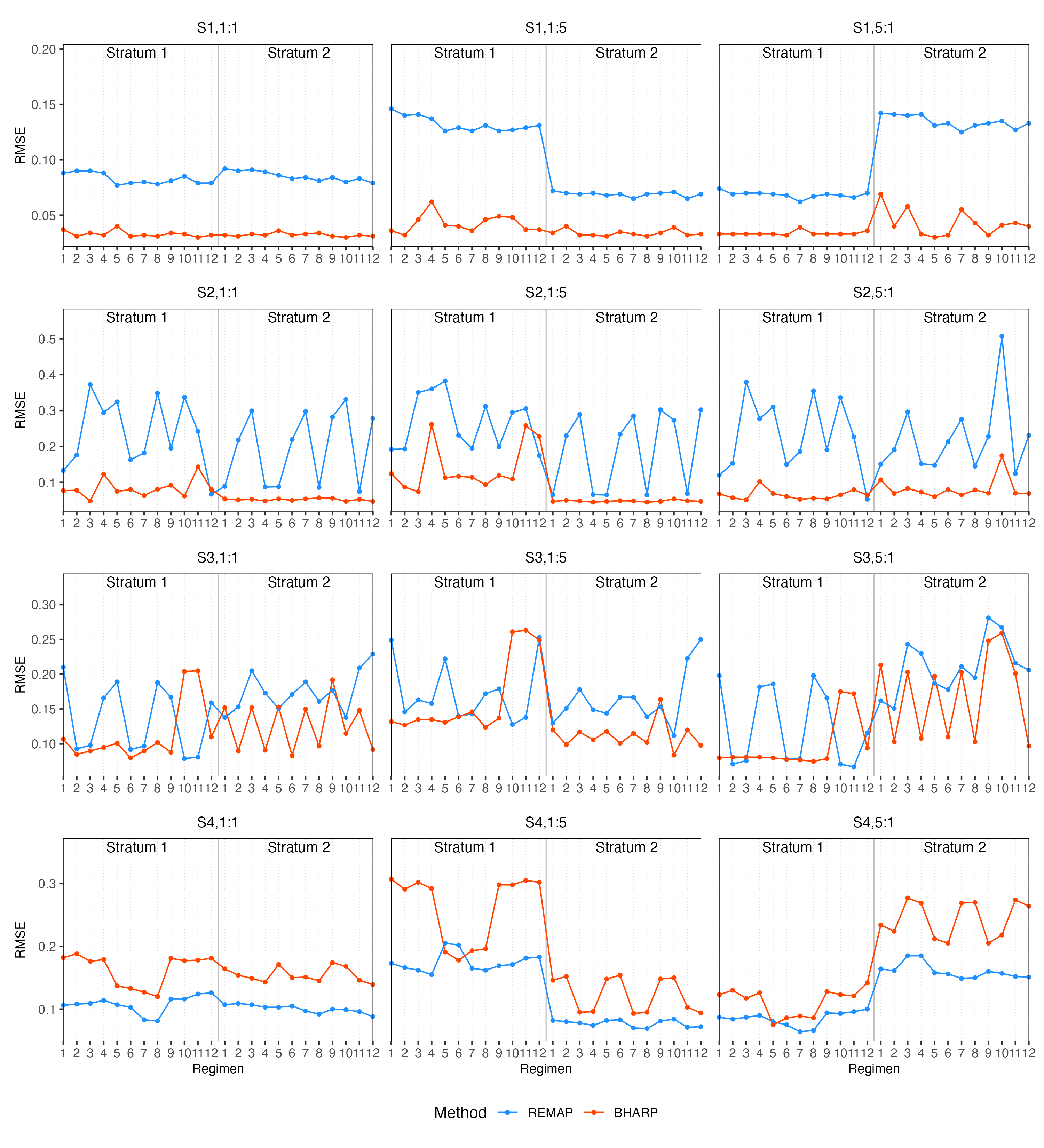}
  \caption{Root mean squared error (RMSE) of cell-mean estimates by scenario, stratum prevalence ratio, stratum, and regimen under the conservative decision thresholds $p_1=0.99$ and $p_2=0.01$.}
  \label{fig:rmse_9901}
\end{figure}

\begin{table}[t]
\centering
\caption{Empirical distribution of trial termination across analysis stages by scenario and stratum prevalence ratio under the conservative decision thresholds $p_1=0.99$ and $p_2=0.01$.}
\label{tab:stage_distribution_9901}
\small
\setlength{\tabcolsep}{4pt}
\begin{tabular}{lcccccc}
\toprule
&  \multicolumn{2}{c}{Stage 1} 
& \multicolumn{2}{c}{Stage 2} 
& \multicolumn{2}{c}{Stage 3} \\
\cmidrule(lr){2-3} \cmidrule(lr){4-5} \cmidrule(lr){6-7}
Scenario 
& REMAP & BHARP 
& REMAP & BHARP 
& REMAP & BHARP \\
\midrule
S1, 1:1 & 0.00 & 0.00 & 0.00 & 0.00 & 1.00 & 1.00 \\
S1, 1:5 & 0.00 & 0.00 & 0.00 & 0.00 & 1.00 & 1.00 \\
S1, 5:1 & 0.00 & 0.00 & 0.00 & 0.00 & 1.00 & 1.00 \\
\midrule
S2, 1:1 & 0.00 & 0.01 & 0.00 & 0.10 & 1.00 & 0.89 \\
S2, 1:5 & 0.00 & 0.00 & 0.00 & 0.03 & 1.00 & 0.97 \\
S2, 5:1 & 0.00 & 0.00 & 0.01 & 0.05 & 0.99 & 0.95 \\
\midrule
S3, 1:1 & 0.00 & 0.00 & 0.00 & 0.00 & 1.00 & 1.00 \\
S3, 1:5 & 0.00 & 0.00 & 0.00 & 0.00 & 1.00 & 1.00 \\
S3, 5:1 & 0.00 & 0.00 & 0.00 & 0.00 & 1.00 & 1.00 \\
\midrule
S4, 1:1 & 0.00 & 0.00 & 0.00 & 0.00 & 1.00 & 1.00 \\
S4, 1:5 & 0.00 & 0.00 & 0.00 & 0.00 & 1.00 & 1.00 \\
S4, 5:1 & 0.00 & 0.00 & 0.00 & 0.00 & 1.00 & 1.00 \\
\bottomrule
\end{tabular}
\end{table}

\begin{table}[t]
\centering
\caption{Empirical probability of incorrect intervention deactivation by scenario, stratum prevalence ratio, stratum, and domain under the conservative decision thresholds $p_1=0.99$ and $p_2=0.01$.}
\label{tab:incorrect_decisions_9901}
\small
\setlength{\tabcolsep}{4pt}
\begin{tabular}{l l ccc ccc c}
\toprule
&  & \multicolumn{3}{c}{$k=1$} 
& \multicolumn{3}{c}{$k=2$} 
& \\
\cmidrule(lr){3-5} \cmidrule(lr){6-8}
Scenario 
& Method
& $d=1$ & $d=2$ & $d=3$
& $d=1$ & $d=2$ & $d=3$
& Trial total \\
\midrule
S1, 1:1 & 
REMAP & 0.04 & 0.02 & 0.05 & 0.03 & 0.01 & 0.08 & 0.20 \\
& BHARP & 0.00 & 0.00 & 0.00 & 0.00 & 0.00 & 0.00 & 0.00 \\
S1, 1:5 & 
REMAP & 0.04 & 0.01 & 0.03 & 0.03 & 0.03 & 0.10 & 0.21 \\
& BHARP & 0.00 & 0.00 & 0.01 & 0.00 & 0.00 & 0.00 & 0.01 \\
S1, 5:1 & 
REMAP & 0.02 & 0.02 & 0.09 & 0.03 & 0.02 & 0.02 & 0.19 \\
& BHARP & 0.00 & 0.00 & 0.00 & 0.00 & 0.00 & 0.00 & 0.01 \\
\midrule
S2, 1:1 & 
REMAP & 0.00 & 0.00 & 0.00 & 0.00 & 0.00 & 0.00 & 0.00 \\
& BHARP & 0.00 & 0.00 & 0.00 & 0.00 & 0.00 & 0.00 & 0.00 \\
S2, 1:5 & 
REMAP & 0.01 & 0.00 & 0.00 & 0.00 & 0.00 & 0.00 & 0.01 \\
& BHARP & 0.00 & 0.00 & 0.00 & 0.00 & 0.00 & 0.00 & 0.00 \\
S2, 5:1 & 
REMAP & 0.00 & 0.00 & 0.00 & 0.00 & 0.01 & 0.00 & 0.02 \\
& BHARP & 0.00 & 0.00 & 0.00 & 0.00 & 0.01 & 0.00 & 0.01 \\
\midrule
S3, 1:1 & 
REMAP & 0.00 & 0.00 & 0.00 & 0.00 & 0.00 & 0.00 & 0.00 \\
& BHARP & 0.00 & 0.00 & 0.00 & 0.00 & 0.00 & 0.00 & 0.00 \\
S3, 1:5 & 
REMAP & 0.01 & 0.01 & 0.00 & 0.00 & 0.00 & 0.00 & 0.01 \\
& BHARP & 0.01 & 0.00 & 0.00 & 0.00 & 0.00 & 0.00 & 0.01 \\
S3, 5:1 & 
REMAP & 0.00 & 0.00 & 0.00 & 0.00 & 0.00 & 0.00 & 0.00 \\
& BHARP & 0.00 & 0.00 & 0.00 & 0.00 & 0.00 & 0.00 & 0.00 \\
\midrule
S4, 1:1 & 
REMAP & 0.04 & 0.05 & 0.00 & 0.03 & 0.00 & 0.15 & 0.26 \\
& BHARP & 0.05 & 0.05 & 0.00 & 0.04 & 0.00 & 0.15 & 0.28 \\
S4, 1:5 & 
REMAP & 0.04 & 0.05 & 0.00 & 0.04 & 0.00 & 0.13 & 0.26 \\
& BHARP & 0.05 & 0.05 & 0.00 & 0.04 & 0.00 & 0.14 & 0.25 \\
S4, 5:1 & 
REMAP & 0.04 & 0.06 & 0.00 & 0.03 & 0.00 & 0.09 & 0.21 \\
& BHARP & 0.04 & 0.04 & 0.00 & 0.05 & 0.00 & 0.14 & 0.24 \\

\bottomrule
\end{tabular}

\begin{minipage}{0.55\linewidth}
\vspace{4pt}
\footnotesize
\textit{Note:} Incorrect deactivation refers to deactivation of at least one true optimal intervention. The ``Trial'' column reports the probability of at least one incorrect deactivation anywhere in the trial.
\end{minipage}

\end{table}

\end{document}